\documentclass[a4paper,fleqn]{cas-sc}

\usepackage[authoryear,longnamesfirst]{natbib}
\usepackage{amsthm,amsmath,amsfonts}
\usepackage{graphicx}
\usepackage{url}
\usepackage{color}
\usepackage{hyperref}
\usepackage{booktabs}

\begin{document}
\let\WriteBookmarks\relax
\def\floatpagepagefraction{1}
\def\textpagefraction{.001}

% Short title
\shorttitle{Hierarchical Spline-Based Bayesian Beta-Binomial Regression}

% Short authors
\shortauthors{Jacobs, Piburn, and Myers}

% Main title
\title[mode = title]{Hierarchical Spline-Based Bayesian Beta-Binomial Regression for Estimating Time-Varying Risk in Power Outages}

% Title footnote mark for DOE notice
\tnotemark[1]
\tnotetext[1]{This manuscript has been authored by UT-Battelle, LLC, under contract DE-AC05-00OR22725 with the US Department of Energy (DOE). The US government retains and the publisher, by accepting the article for publication, acknowledges that the US government retains a nonexclusive, paid-up, irrevocable, worldwide license to publish or reproduce the published form of this manuscript, or allow others to do so, for US government purposes. DOE will provide public access to these results of federally sponsored research in accordance with the DOE Public Access Plan (\url{https://www.energy.gov/doe-public-access-plan}).}

% First author
\author[1]{Justin Jacobs}
\cormark[1]
\fnmark[1]
\ead{jacobsjw@ornl.gov}
\credit{Conceptualization, Methodology, Software, Writing -- Original Draft}

% Second author
\author[1]{Jesse Piburn}
\fnmark[1]
\ead{piburnjo@ornl.gov}
\credit{Methodology, Software, Writing -- Review \& Editing}

% Third author
\author[1]{Aaron Myers}
\fnmark[1]
\ead{myersat@ornl.gov}
\credit{Methodology, Writing -- Review \& Editing}

% Affiliation
\affiliation[1]{organization={Oak Ridge National Laboratory},
            addressline={1 Bethel Valley Road},
            city={Oak Ridge},
            postcode={37831},
            state={TN},
            country={USA}}

% Corresponding author text
\cortext[1]{Corresponding author}

% Footnote text
\fntext[1]{Oak Ridge National Laboratory}

% Abstract
\begin{abstract}
We propose a hierarchical Bayesian model to infer time-varying risk of power outages across multiple geographic groups. Using cubic B-spline basis functions and a beta-binomial likelihood, we estimate latent outage probabilities over time while accounting for overdispersion. Posterior inference is conducted via MCMC in PyMC, and we assess predictive performance using RMSE, credible intervals, and area under the curve (AUC) metrics. The model is applied to power outage data, demonstrating its effectiveness in capturing dynamic risk across diverse settings.
\end{abstract}

% Research highlights
\begin{highlights}
\item Hierarchical Bayesian B-spline model estimates time-varying power outage risk with full uncertainty quantification.
\item Beta-Binomial likelihood with hierarchical dispersion pooling accommodates overdispersion across geographic groups.
\item Posterior AUC distributions provide reliability-aware resilience metrics unavailable from deterministic methods.
\end{highlights}

% Keywords

\vspace{1em}
  \noindent\textbf{Notice of Copyright:}
  \small
  \textit{This manuscript has been authored by UT-Battelle, LLC, under contract DE-AC05-00OR22725 with the US Department of Energy (DOE). The US government retains and the publisher, by accepting the article for publication, acknowledges that the US government retains a nonexclusive, paid-up, irrevocable, worldwide license to publish or reproduce the published form of this manuscript, or allow others to do so, for US government purposes. DOE will provide public access to these results of federally sponsored research in accordance with the DOE Public Access Plan (\href{https://www.energy.gov/doe-public-access-plan}{https://www.energy.gov/doe-public-access-plan}).}  % Full block here

\maketitle

%%%%%%%%%%%%%%%%%%%%%%%%%%%%%%%%%%%%%%%%%%%%%%%%%%%%%%%%%%%%%
\section{Introduction}

Extreme weather events and infrastructure failures continue to pose significant risks to electric power systems, resulting in widespread outages that disrupt essential services and incur substantial economic costs \citep{lee2024quantifying}. These risks vary geographically and temporally, with localized factors---such as vegetation, equipment age, and weather intensity---shaping the severity and recovery trajectory of outages. To understand and mitigate these risks, recent work has emphasized the need for high-resolution spatiotemporal models of outage impacts that can support resilience planning and decision-making under uncertainty.

A substantial body of work has developed formal frameworks and metrics to quantify power system resilience. Panteli et al.\ \cite{panteli2017metrics} introduced the resilience trapezoid, decomposing system performance during an extreme event into three phases: disturbance progress, post-disturbance degraded state, and recovery. Their $\Phi\Lambda\text{E}\Pi$ metric system characterizes how fast and how low resilience drops, how long the degraded state persists, and how promptly the system recovers, with the area of the trapezoid serving as a complementary scalar summary of cumulative impact. Francis and Bekera \cite{francis2014metric} proposed a complementary resilience factor synthesizing absorptive, adaptive, and restorative capacities under uncertainty, demonstrating its application to electric power distribution networks subject to hurricane hazards. Yao et al. \cite{yao2022framework} reviewed and categorized existing resilience metrics into pre-event estimation and post-event evaluation frameworks, noting that while post-event metrics are well established, comparability across systems and events remains a challenge. Across these frameworks, a common limitation emerges: resilience metrics are computed from deterministic performance trajectories, yielding scalar summaries with no attached uncertainty. As a result, two outage events with identical trapezoid areas but very different data densities are treated as statistically equivalent, masking meaningful differences in inferential confidence.

Recent efforts to characterize power system resilience have largely focused on identifying outage events and summarizing their dynamics through scalar metrics such as impact duration, recovery time, and customer exposure \citep{lee2024comprehensive, nateghi2011comparison}. A prominent example is \citet{lee2024comprehensive}, who proposed a parameterized piecewise curve model to represent the escalation and recovery phases of outages. While interpretable, this approach requires manual breakpoint selection and yields a single deterministic trajectory per event---producing no uncertainty estimates over derived metrics such as AUC. Similarly, the resilience framework of \citet{lee2024quantifying} computes cumulative outage metrics such as AUC and lagged recovery directly from raw time series, without accounting for observation noise, reporting irregularities, or the overdispersion commonly present in county-level outage counts.

Parallel to these descriptive efforts, probabilistic outage modeling approaches fall broadly into two categories: predictive models that forecast outage counts as a function of weather and infrastructure covariates, and hierarchical models that estimate recovery or duration parameters across geographic groups. In the first category, spline-based regression has been applied to outage risk primarily through generalized additive models (GAMs) \citep{wood2017generalized}, which replace the linear predictor of a GLM with flexible smooth functions of covariates estimated nonparametrically. Applied to power systems, GAMs have demonstrated substantially improved predictive accuracy over GLMs by capturing nonlinear relationships between wind speed, soil moisture, and vegetation conditions and outage counts \citep{han2009improving, nateghi2014power}, while ensemble methods such as Random Forests \citep{tervo2019short} and hybrid classification-GAM approaches \citep{guikema2014predictive} have further extended predictive performance. These approaches are designed for cross-sectional prediction of outage counts as a function of pre-event covariates, however, and are not structured to produce posterior distributions over the full temporal trajectory of an unfolding event. In the second category, \citet{rao2025power} developed hierarchical Bayesian models using a gamma likelihood to estimate average outage durations across counties in the U.S.\ Gulf Coast, incorporating social vulnerability and hazard type as covariates. Similarly, \citet{owolabi2022bayesian} applied hierarchical time-series forecasting with Bayesian hyperparameter optimization to model non-weather-related outage counts at the county level. These approaches can achieve strong predictive accuracy or recover meaningful group-level structure, but they are not designed to produce posterior distributions over the full temporal trajectory of a single outage event. The former class models outage counts as a function of pre-event covariates; the latter estimates scalar duration summaries or aggregate counts across periods. Neither class yields a posterior distribution over the latent risk curve from which derived resilience metrics such as AUC can be computed with calibrated uncertainty. The absence of formal uncertainty quantification over the recovery trajectory is a structural limitation of these frameworks, not an implementation gap that can be patched by adding confidence intervals post hoc.

To address these limitations, hierarchical Bayesian models have emerged as a principled framework for inferring latent structure while accounting for group-level heterogeneity. Within the Bayesian hierarchical literature for power systems, existing work has focused primarily on static reliability parameters. \citet{zhou2020hierarchical} proposed a Bayesian hierarchical model leveraging spatial and network proximity kernels to estimate individual transmission line outage rates, demonstrating that partial pooling across lines substantially reduces estimation uncertainty relative to conventional frequency-based methods. \citet{dunn2020regional} developed a related framework for estimating component-level failure probabilities from fragility curves under wind loading, showing that hierarchical pooling over heterogeneous components improves both parameter recovery and upgrade policy recommendations. \citet{iesmantas2014bayesian} applied a Bayesian hierarchical model with a Borel-Tanner distribution to cascading outage data from North American transmission lines, recovering evidence of non-negligible large-blackout probabilities that simpler Poisson models underestimate. While these frameworks demonstrate the value of hierarchical Bayesian inference for power system reliability, they estimate static component outage rates or aggregate cascade statistics rather than the temporal trajectory of an unfolding event. Consequently, none yield posterior distributions over time-varying risk curves from which derived resilience metrics such as AUC, peak severity, or recovery rate can be computed with calibrated uncertainty. A key advantage of the Bayesian paradigm in this context is that uncertainty propagates through the full model---from spline coefficients through the sigmoid transformation and into derived quantities such as AUC---yielding credible intervals that quantify what we do not know about the recovery trajectory itself.

A parallel body of work has developed spatiotemporal frameworks that propagate outage risk across geographic units during evolving hazard events. \citet{xu2024hazard} proposed a hazard resistance-based spatiotemporal risk analysis method for distribution networks during hurricanes, converting time-varying failure probabilities into time-invariant hazard resistances for each network component and validating the approach against observed outage data from Puerto Rico during Hurricane Fiona. \citet{jiang2024spatiotemporal} developed a graph conformal prediction framework that models spatio-temporal dependencies in customer-level outage counts using a Poisson regression base model and quantile random forests for uncertainty quantification across geographically connected units. These frameworks capture spatial correlation in outage propagation across the power network, but they operate at the aggregate prediction level and do not model the full posterior distribution over the temporal trajectory of a single event's recovery process. The present work focuses on the latter problem---estimating time-varying latent risk within individual outage events from observed customer impact data---and treats geographic groups as exchangeable given the hierarchical prior on dispersion. Spatial extensions, such as adjacency-aware pooling or CAR priors over county-level trajectories, are deferred to future work.

Despite the breadth of existing approaches, a specific inferential gap remains unaddressed. Deterministic resilience frameworks \citep{panteli2017metrics, francis2014metric, lee2024quantifying} produce scalar summaries from observed trajectories without attaching uncertainty to those summaries, treating events with three observations and three hundred as equally well-characterized. Predictive outage models---whether GAM-based \citep{han2009improving, nateghi2014power} or ensemble-based \citep{tervo2019short, guikema2014predictive}---are designed for cross-sectional forecasting from pre-event covariates and are not structured to produce posterior distributions over the temporal trajectory of an unfolding event. Hierarchical Bayesian frameworks for power systems \citep{zhou2020hierarchical, dunn2020regional, iesmantas2014bayesian} estimate static component reliability parameters or aggregate cascade statistics rather than time-varying latent risk curves. Spatiotemporal frameworks \citep{xu2024hazard, jiang2024spatiotemporal} propagate risk across network components but operate at the aggregate prediction level and do not yield posterior distributions over the recovery trajectory of a single event. Hierarchical Bayesian outage duration models \citep{rao2025power, owolabi2022bayesian} recover group-level scalar summaries but do not model continuous risk trajectories from which derived metrics such as AUC can be computed with calibrated uncertainty. No existing framework simultaneously models the full temporal trajectory of an individual outage event, accommodates overdispersion in county-level counts through a flexible likelihood, and propagates uncertainty through to derived resilience metrics such as AUC. The present work fills this gap.

Within the Bayesian paradigm, spline-based regression models offer a principled alternative to frequentist GAMs: rather than producing point estimates with asymptotic confidence intervals, they yield full posterior distributions over smooth latent functions, enabling uncertainty propagation through derived quantities such as AUC, peak severity, and recovery rate \citep{crainiceanu2005penalized}. Fully Bayesian extensions of penalized splines \citep{fahrmeir2001bayesian, reich2010bayesian} have demonstrated flexible trajectory estimation under uncertainty across environmental and biomedical domains. This distinction is consequential in the outage risk context, where data sparsity and reporting irregularities make point-estimate smoothing unreliable and where downstream resilience decisions depend on knowing how much confidence the inferred trajectory warrants. Despite this potential, Bayesian spline regression has not been applied to the event-level outage risk problem addressed here.

In this paper, we introduce a hierarchical Bayesian model that estimates time-varying outage risk by combining cubic B-spline regression with a Beta-Binomial likelihood. The latent risk curve for each region is modeled using spline basis functions, allowing for smooth trajectories that capture nonlinear variation over time. To account for overdispersion in the outage counts, we incorporate a shared hyperprior on the concentration parameter $\kappa_{\text{global}}$, enabling hierarchical shrinkage and improved estimation across geographically indexed groups. This formulation yields full posterior distributions over the underlying outage probabilities, from which we derive credible intervals and summary measures such as the area under the risk curve (AUC). Our implementation supports posterior predictive checks, root mean square error (RMSE) comparisons, and leave-one-out (LOO) cross-validation, providing both interpretability and rigorous uncertainty quantification \citep{gelman2020bayesian}.

By embedding our model within a Bayesian spline regression framework, we offer a statistically grounded alternative to deterministic curve-fitting approaches. This enables principled uncertainty quantification, posterior simulation, and comparative risk assessment across regions—making it a flexible tool for resilience planning and inference from sparse, heterogeneous outage data.

%%%%%%%%%%%%%%%%%%%%%%%%%%%%%%%%%%%%%%%%%%%%%%%%%%%%%%%%%%%%%
\section{Model Specification}

To characterize power system resilience through observed outage events, we adopt a Bayesian hierarchical model that fits B-splines to the log-odds of outage probability. This approach allows us to capture the evolving \emph{velocity} and \emph{acceleration} of recovery efforts by leveraging the first and second derivatives of the spline basis \cite{ramsay2005functional, crainiceanu2005penalized}. The resulting regression defines a latent trajectory of outage risk throughout each event and enables derived metrics, such as area under the curve (AUC), to be estimated within a fully probabilistic framework.

As outage dynamics vary across geographic locations and storm conditions, our model introduces partial pooling to account for cross-group heterogeneity while borrowing strength across events. Due to differences in infrastructure, reporting fidelity, or damage severity, overdispersion in observed counts is accommodated through a flexible Beta-Binomial likelihood. The combination of spline smoothing and hierarchical priors yields a robust posterior distribution over group-specific outage trajectories, from which we can derive credible intervals, uncertainty estimates, and summary risk measures.

\subsection{EAGLE-I Data and Event-Based Observational Framework}

The \textit{Environment for Analysis of Geo-Located Energy Information (EAGLE-I\texttrademark)} is a geospatial monitoring platform developed by the U.S. Department of Energy and maintained by Oak Ridge National Laboratory. It provides the federal government’s authoritative, near real-time situational awareness for monitoring energy infrastructure across the electricity, natural gas, and petroleum sectors. EAGLE-I integrates operational data with interactive geographic overlays, including utility outage information, critical infrastructure locations, and weather impacts, to support coordinated emergency response \cite{Brelsford2024eaglei}.

For each county in the United States, EAGLE-I aggregates and updates power outage data approximately every 15 minutes, reporting:
\begin{itemize}
    \item Number of customers currently without power,
    \item Total customer base,
    \item Time of observation, and
    \item County identifier (via FIPS code).
\end{itemize}

Data is recorded only when at least one customer is experiencing an outage, producing a zero-truncated time series. While this sparsity poses challenges for continuous-time modeling, we focus our analysis on discrete \emph{outage events}.

Following prior resilience studies \cite{lee2024comprehensive, lee2024quantifying}, we define an outage event as a window of time beginning when the number of affected customers exceeds a predefined threshold. We model only the duration of disruption and recovery, with the time index re-centered for each group such that \( t = 0 \) marks the start of the event. Observed quantities include:
\begin{itemize}
    \item \( y_{g,t} \): number of customers without power in group \( g \) at time \( t \),
    \item \( n_{g,t} \): total number of customers in the group, optionally scaled using the national average EAGLE-I coverage ratio (\( \approx 0.871 \)).
\end{itemize}

We illustrate this data structure using an EAGLE-I extract from Dane County, Wisconsin, on May 22, 2024. That evening, severe thunderstorms swept through southern Wisconsin, toppling infrastructure and leaving over 40,000 customers without power across multiple counties. Madison Gas \& Electric (MG\&E), We Energies, and Alliant Energy reported widespread disruptions, with MG\&E calling it one of the largest outage events in decades. By 5:00 p.m., over 8,300 MG\&E customers remained without power, with thousands more affected in neighboring counties.

\begin{figure}[htbp]
    \centering
    \includegraphics[width=0.9\textwidth]{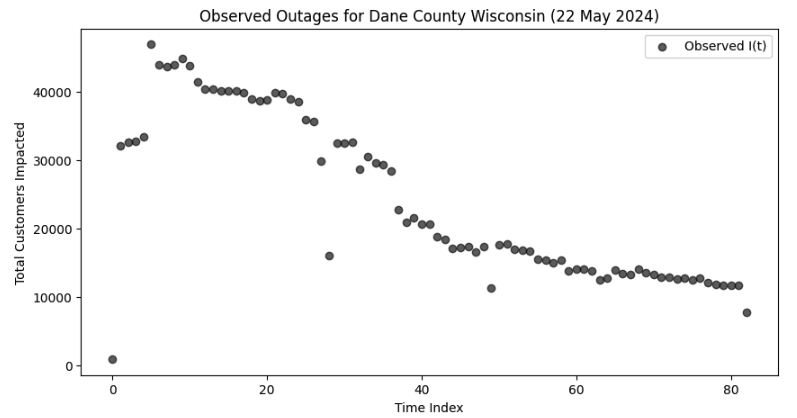}
    \caption{Observed power outages in Dane County, Wisconsin, on May 22, 2024, based on EAGLE-I event-based data. A severe thunderstorm led to a sharp increase in outages, peaking above 40,000 customers before a gradual recovery. Data shown reflects total customers impacted at each 15-minute interval after the threshold was exceeded.}
    \label{fig:dane_event}
\end{figure}

This example illustrates the type of localized, high-resolution data captured by EAGLE-I and motivates the use of probabilistic models that can flexibly represent the shape and uncertainty of recovery trajectories under diverse operational conditions.

We specify the following likelihood for the observed number of customers without power:
\begin{equation}
\label{eq:beta_binomial_likelihood}
y_{g,t} \sim \text{BetaBinomial}(n_{g,t}, \alpha_{g,t}, \eta_{g,t}), \quad
\text{where} \quad 
\alpha_{g,t} = p_{g,t} \cdot \kappa_g, \quad
\eta_{g,t} = (1 - p_{g,t}) \cdot \kappa_g.
\end{equation}
where \( p_{g,t} \in (0,1) \) is the latent outage probability and \( \kappa_g > 1 \) governs overdispersion within group \( g \). This model enables full posterior inference over the risk curve for each outage event and supports the construction of credible bands, event-level summaries, and probabilistic comparisons across space and time.

The Beta-Binomial likelihood is particularly well suited for this application as it generalizes the Binomial model by allowing the underlying event probability \( p_{g,t} \) to vary according to a Beta distribution. The concentration parameter \( \kappa_g \) governs the dispersion of this distribution: as \( \kappa_g \to \infty \), the model reduces to a standard Binomial likelihood with fixed probability \( p_{g,t} \); when \( \kappa_g \) is small, greater uncertainty is allowed in the underlying probability, reflecting variability due to unobserved heterogeneity such as infrastructure conditions, weather severity, or reporting reliability.

By estimating \( \kappa_g \) hierarchically across groups, we allow the degree of overdispersion to adapt to each event while borrowing strength from the overall population. This setup is especially valuable in cases where some outage events are sparse or short-lived, as it allows the model to avoid overfitting group-specific noise while still capturing meaningful variation in recovery dynamics.

In the Bayesian setting, this likelihood structure allows us to obtain posterior distributions not only over the latent probabilities \( p_{g,t} \), but also over derived functionals such as the total impacted duration, peak severity, or cumulative risk (e.g., AUC). These summaries are central to assessing power system resilience and quantifying recovery efficiency across diverse spatial and temporal contexts.

\subsection{Prior Specification}

From the likelihood in Equation~\eqref{eq:beta_binomial_likelihood}, we model the latent outage probability \( p_{g,t} \) as a sigmoid-transformed linear combination of B-spline basis functions:
\begin{equation}
\label{eq:sigmoid_transform}
p_{g,t} = \sigma(f_g(t)) = \frac{1}{1 + e^{-f_g(t)}}, \quad \text{where} \quad f_g(t) = \mathbf{B}(t)^\top \boldsymbol{\beta}_g.
\end{equation}
This transformation maps the unbounded latent function \( f_g(t) \) to the unit interval, ensuring valid probability estimates. The B-spline basis \( \mathbf{B}(t) \) provides a smooth and flexible representation of the time-varying log-odds of outage risk within each event.

To regularize the spline-based regression and prevent overfitting, we place a weakly informative Gaussian prior on the spline coefficients:
\begin{equation}
\label{eq:beta_prior}
\boldsymbol{\beta}_g \sim \mathcal{N}(0, \tau^2 I),
\end{equation}
where \( \tau^2 \) controls the prior variance and thus the overall smoothness of \( f_g(t) \) in \eqref{eq:sigmoid_transform}. This prior does not induce partial pooling across groups, but it acts as a form of shrinkage within each event, constraining the latent trajectory to a plausible scale while retaining flexibility.

Partial pooling in our model is introduced through a hierarchical prior on the Beta-Binomial concentration parameter \( \kappa_g \), which controls the degree of overdispersion in each group’s observed outage counts (see Equation~\eqref{eq:beta_binomial_likelihood}). While the Gaussian prior on the spline coefficients \( \boldsymbol{\beta}_g \) in Equation~\eqref{eq:beta_prior} regularizes the scale and smoothness of the latent risk trajectory within each event, it does not induce information sharing across groups.

To introduce structured sharing, we place a hyperprior on a global concentration parameter \( \kappa_{\text{global}} \), which governs the distribution of group-specific dispersion parameters. Each group's overdispersion parameter \( \kappa_g \) is defined by adding one to a latent variable \( \kappa_{\text{raw},g} \), which is drawn from a Gamma distribution centered around \( \kappa_{\text{global}} \). This hierarchical structure enables groups with sparse or noisy data to borrow statistical strength from the population while allowing sufficient flexibility for well-informed groups to express distinct behavior.

Formally, we specify:
\begin{equation}
\label{eq:kappa_hierarchy}
\kappa_{\text{global}} \sim \text{Gamma}(5, 0.5), \qquad
\kappa_g = \kappa_{\text{raw},g} + 1, \qquad
\kappa_{\text{raw},g} \sim \text{Gamma}(5, 5 / \kappa_{\text{global}}).
\end{equation}

Adding one ensures that \( \kappa_g > 1 \), which preserves desirable properties for the Beta distribution in our likelihood and avoids degenerate behavior associated with low-concentration regimes. This formulation reflects our belief that while overdispersion varies across events, it does so in a structured and learnable manner—critical for modeling power outage dynamics at scale.

\subsection{Full Model Specification}

The full generative model combines the spline-based regression for latent outage risk with a Beta-Binomial observation model and a hierarchical prior on group-level dispersion. For each group \( g \) and time index \( t \), the model proceeds as follows:

\begin{enumerate}
    \item \textbf{Spline-based latent function:}
    \[
    f_g(t) = \mathbf{B}(t)^\top \boldsymbol{\beta}_g
    \]
    where \( \mathbf{B}(t) \) is the vector of cubic B-spline basis functions evaluated at time \( t \), and \( \boldsymbol{\beta}_g \sim \mathcal{N}(0, \tau^2 I) \) is a vector of spline coefficients with a weakly informative Gaussian prior.

    \item \textbf{Latent probability via sigmoid transform:}
    \[
    p_{g,t} = \sigma(f_g(t)) = \frac{1}{1 + \exp(-f_g(t))}
    \]

    \item \textbf{Hierarchical prior on dispersion:}
    \begin{align}
    \kappa_{\text{global}} &\sim \text{Gamma}(5, 0.5) \\
    \kappa_{\text{raw},g} &\sim \text{Gamma}(5, 5 / \kappa_{\text{global}}) \\
    \kappa_g &= \kappa_{\text{raw},g} + 1
    \end{align}

    \item \textbf{Beta-Binomial likelihood:}
    \begin{align}
    \alpha_{g,t} &= p_{g,t} \cdot \kappa_g \\
    \eta_{g,t} &= (1 - p_{g,t}) \cdot \kappa_g \\
    y_{g,t} &\sim \text{BetaBinomial}(n_{g,t}, \alpha_{g,t}, \eta_{g,t})
    \end{align}
\end{enumerate}

This formulation supports flexible, event-specific modeling of outage risk trajectories while introducing structured regularization through both Gaussian priors (on spline coefficients) and hierarchical pooling (on dispersion parameters). The resulting generative model defines a joint posterior distribution over all parameters, which we target using MCMC-based inference methods described in the next section. In addition to posterior distributions over latent functions and dispersion terms, this framework enables posterior predictive inference for the observed outages \( y_{g,t} \). These predictive quantities form the basis for assessing model fit, constructing credible intervals, and quantifying event-level metrics such as peak load and area under the risk curve (AUC).

%%%%%%%%%%%%%%%%%%%%%%%%%%%%%%%%%%%%%%%%%%%%%%%%%%%%%%%%%%%%%
\section{Posterior Inference}

Given the complexity of the hierarchical model described in the Full Model Specification, we perform full Bayesian inference using Markov Chain Monte Carlo (MCMC) sampling. This approach allows us to draw from the joint posterior distribution over spline coefficients, dispersion parameters, and derived quantities such as event-level risk trajectories. Posterior samples enable a range of downstream analyses, including uncertainty quantification, predictive validation, and estimation of interpretable summaries such as peak outage levels or cumulative risk. We describe the posterior structure formally below, followed by methods for posterior predictive simulation and evaluation.

\subsection{Posterior Distribution}

Let \( \mathcal{D} = \{y_{g,t}, n_{g,t}\}_{g,t} \) denote the observed outage data across all groups, \(g\), and time points, \(t\). The joint posterior distribution over the model parameters is given by:

\begin{equation}
\label{eq:posterior}
p\left(\{\boldsymbol{\beta}_g\}, \{\kappa_g\}, \kappa_{\text{global}} \mid \mathcal{D}\right)
\propto
p(\mathcal{D} \mid \{\boldsymbol{\beta}_g\}, \{\kappa_g\}) \cdot
p(\{\boldsymbol{\beta}_g\}) \cdot
p(\{\kappa_g\} \mid \kappa_{\text{global}}) \cdot
p(\kappa_{\text{global}}).
\end{equation}

\noindent where:
\begin{itemize}
  \item \( p(\mathcal{D} \mid \{\boldsymbol{\beta}_g\}, \{\kappa_g\}) \) is the Beta-Binomial likelihood based on latent probabilities \( p_{g,t} = \sigma(f_g(t)) \),
  \item \( p(\{\boldsymbol{\beta}_g\}) = \prod_g \mathcal{N}(\boldsymbol{\beta}_g \mid \mathbf{0}, \tau^2 I) \) regularizes the spline coefficients,
  \item \( p(\{\kappa_g\} \mid \kappa_{\text{global}}) = \prod_g \text{Gamma}(\kappa_g - 1 \mid 5, 5 / \kappa_{\text{global}}) \) induces partial pooling across groups,
  \item \( p(\kappa_{\text{global}}) = \text{Gamma}(5, 0.5) \) serves as a shared hyperprior.
\end{itemize}

Inference is carried out using the No-U-Turn Sampler (NUTS), a gradient-based MCMC algorithm implemented in PyMC. We run two chains with 1,000 warm-up steps and 1,000 posterior samples each. The resulting draws provide approximate samples from the posterior distribution over all latent and observed quantities. These samples support posterior summaries such as credible intervals for outage risk \( p_{g,t} \), predictive simulations for impacted customers \( y_{g,t} \), and functionals such as area under the curve (AUC), discussed below.

\subsection{Area Under the Curve (AUC)}

The area under the curve (AUC) of an outage risk trajectory provides an interpretable summary of both the duration and severity of a disruption. In the context of power system resilience, a larger AUC reflects more sustained or widespread outages, while a smaller AUC implies faster recovery or fewer customers affected. As such, AUC serves as a useful scalar summary of event-level impact, enabling direct comparison across outages of varying intensity and length.

Prior work has identified area-based metrics as critical summaries of resilience performance. Panteli et al.\ \cite{panteli2017metrics} introduced the trapezoid area as a scalar metric encoding both the severity and duration of degradation across all phases of an outage event, while Francis and Bekera \cite{francis2014metric} similarly emphasized integral-based performance measures for characterizing absorptive and restorative capacity. In the context of observed outage data, Lee et al.\ \cite{lee2024quantifying} evaluated cumulative outage impact across spatially distributed storm events, using AUC to compare regional performance and recovery dynamics. Similarly, \cite{carrington2021extracting}, defines customer-hours not served as the integral of the outage trajectory over time, we compute the area under the inferred probability curve \( p_{g,t} \) to characterize the scale and duration of each outage event. Unlike the raw outage time series used in \cite{carrington2021extracting}, our AUC metric is derived from a posterior distribution and reflects uncertainty in the underlying risk process.

In our framework, we estimate AUC directly from the posterior distribution of the latent outage probability trajectory \( p_{g,t} \). For each posterior draw of the spline coefficients \( \boldsymbol{\beta}_g \), we evaluate the transformed function \( p_{g,t} = \sigma(f_g(t)) \) across a dense time grid and compute AUC using Simpson’s Rule:

\begin{equation}
\label{eq:auc_simpson}
\text{AUC}_g^{(s)} = \frac{\Delta t}{3} \left[ p_{g,0}^{(s)} + 4 \sum_{\substack{j=1 \\ j \text{ odd}}}^{J-1} p_{g,j}^{(s)} + 2 \sum_{\substack{j=2 \\ j \text{ even}}}^{J-2} p_{g,j}^{(s)} + p_{g,J}^{(s)} \right]
\end{equation}

where \( p_{g,j}^{(s)} \) is the sigmoid-transformed risk at grid point \( j \) for posterior sample \( s \), \( J \) is an even number of subintervals, and \( \Delta t \) is the grid spacing. This yields a posterior distribution over AUC for each group, from which we can compute summary statistics and credible intervals to support resilience analysis.

For each group $g$, we compute the posterior distribution of AUC by applying Equation~\eqref{eq:auc_simpson} to each posterior sample of $\boldsymbol{\beta}_g$. This results in a set of draws $\{\text{AUC}_g^{(s)}\}_{s=1}^S$, where $S$ is the  number of retained MCMC samples. From these draws, we summarize the posterior AUC using the posterior mean $\mathbb{E}[\text{AUC}_g]$, standard deviation, and central credible intervals (e.g., 95\%). These summaries provide not only a point estimate of event impact, but also quantify uncertainty due to model structure, spline flexibility, and data sparsity in a coherent, jointly calibrated way. Critically, this uncertainty characterization is not available from deterministic approaches that integrate raw observed proportions directly: such methods yield a scalar AUC value with no attached measure of reliability, and thus cannot distinguish a well-characterized estimate from 
a highly uncertain one arising from sparse or noisy data. In practice, we report the posterior mean and 2.5th and 97.5th percentiles of $\text{AUC}_g$, which enable consistent and uncertainty-aware comparison of outage dynamics across space and time.

\subsection{Illustrative Example: Dane County, WI}

To demonstrate the application of our model, we focus on an outage event in Dane County, Wisconsin (Group ID 66806), which experienced a major disruption on May 22, 2024, due to severe thunderstorms. As shown in Figure~\ref{fig:dane_event}, over 40{,}000 customers were affected at the peak, with recovery extending over several hours.

Posterior inference for this event was performed using the hierarchical Beta-Binomial model with a seven-dimensional cubic B-spline basis. The basis dimension was selected adaptively using the rule
\[
\texttt{df\_value} = \max\left( \min\left( \max(5, \lfloor n/5 \rfloor), n - 4, 20 \right), 6 \right),
\]
where \( n \) is the number of observed time points. This rule balances three competing goals: (i) ensuring a minimum number of basis functions for cubic B-spline identifiability (degree 3 requires at least 4 basis terms), (ii) scaling model complexity based on data availability (via \( n/5 \)), and (iii) constraining overparameterization in long time series through a maximum df cap of 20. The heuristic choice of \( \lfloor n/5 \rfloor \) is a common default in spline regression, echoing recommendations from penalized spline literature \cite{rupert2003semiparametric}.

The resulting posterior summaries of the spline coefficients \( \beta_0,\dots, \beta_{16} \), shown in Table~\ref{tab:dane_beta_summary}, describe the latent function \( f_{66806}(t) \), which governs the time-varying outage probability via the sigmoid transformation.

\begin{table}[htbp]
\centering
\begin{tabular}{lrrrrr}
\toprule
Parameter & Mean & SD & 2.5\% HDI & 97.5\% HDI & $\hat{R}$ \\
\midrule
$\beta_0$ & -3.726 & 0.426 & -4.603 & -3.001 & 1.00 \\
$\beta_1$ & 2.309 & 0.722 & 0.944 & 3.688 & 1.00 \\
$\beta_2$ & 1.982 & 0.589 & 0.978 & 3.129 & 1.00 \\
$\beta_3$ & 2.156 & 0.566 & 1.036 & 3.180 & 1.00 \\
$\beta_4$ & 1.915 & 0.533 & 0.947 & 2.958 & 1.00 \\
$\beta_5$ & 2.093 & 0.556 & 1.099 & 3.191 & 1.00 \\
$\beta_6$ & 1.437 & 0.548 & 0.413 & 2.449 & 1.00 \\
$\beta_7$ & 1.857 & 0.577 & 0.844 & 3.004 & 1.00 \\
$\beta_8$ & 1.100 & 0.590 & 0.065 & 2.321 & 1.00 \\
$\beta_9$ & 1.073 & 0.608 & -0.121 & 2.156 & 1.00 \\
$\beta_{10}$ & 1.109 & 0.625 & -0.101 & 2.229 & 1.00 \\
$\beta_{11}$ & 0.965 & 0.619 & -0.268 & 2.059 & 1.00 \\
$\beta_{12}$ & 0.829 & 0.629 & -0.350 & 1.995 & 1.00 \\
$\beta_{13}$ & 0.943 & 0.678 & -0.371 & 2.171 & 1.00 \\
$\beta_{14}$ & 0.713 & 0.801 & -0.751 & 2.248 & 1.00 \\
$\beta_{15}$ & 0.904 & 0.829 & -0.606 & 2.396 & 1.00 \\
$\beta_{16}$ & 0.296 & 0.756 & -1.088 & 1.697 & 1.00 \\
$\kappa_{\text{raw},66806}$ & 62.406 & 10.057 & 41.776 & 79.388 & 1.00 \\
\bottomrule
\end{tabular}
\caption{Posterior summary for spline coefficients and dispersion parameter for the 22 May 2024 Dane County, Wisconsin event.}
\label{tab:dane_beta_summary}
\end{table}

The updated posterior summaries offer a more granular view of the temporal evolution of outage risk. The strongly negative intercept \( \beta_0 = -3.73 \) establishes a low baseline probability, while a sequence of large positive coefficients from \( \beta_1 \) through \( \beta_5 \) signals a sharp and sustained escalation in risk, peaking early in the event. This aligns with the rapid onset of outages observed in Dane County. The mid- to late-stage coefficients (\( \beta_6 \) through \( \beta_{10} \)) remain positive but gradually taper, indicating a slower, extended recovery phase. The final few coefficients (\( \beta_{11} \) through \( \beta_{16} \)) diminish in magnitude and include wider credible intervals that span zero, suggesting a flattening of risk toward the event's end. The posterior mean of the raw dispersion parameter \( \kappa_{\text{raw},66806} = 62.4 \) indicates modest overdispersion, implying that observed outage counts were closely aligned with the latent risk trajectory inferred from the model.

\begin{figure}[htbp]
    \centering
    \includegraphics[width=0.85\textwidth]{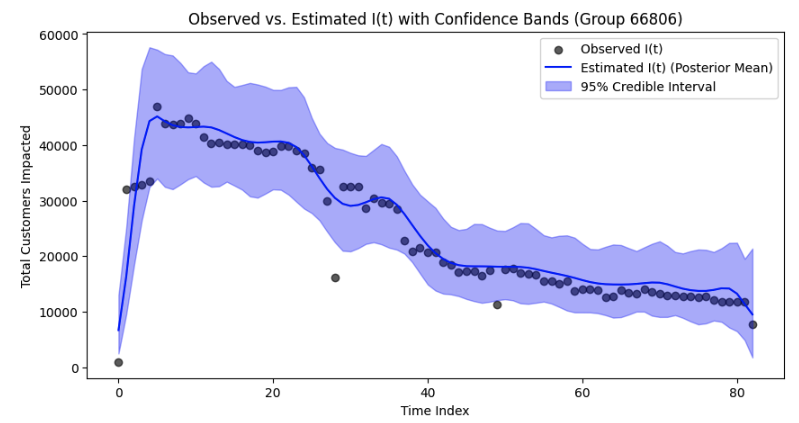}
    \caption{Posterior predictive fit for Dane County, WI (Group ID 66806). The solid blue line represents the posterior mean of estimated outages \( \hat{y}_{g,t} = p_{g,t} \cdot n_{g,t} \), and the shaded region shows the 95\% pointwise credible interval. Black dots represent observed outage counts at 15-minute intervals.}
    \label{fig:dane_posterior_fit}
\end{figure}

Figure~\ref{fig:dane_posterior_fit} shows the posterior predictive fit for the total number of impacted customers \( y_{66806,t} \). The solid line denotes the posterior mean, and the shaded band shows the 95\% pointwise credible interval. Black dots represent the observed data.

We summarize the overall severity of the event using the area under the posterior mean risk curve, estimated via Simpson’s Rule as in Equation~\eqref{eq:auc_simpson}. The posterior distribution of AUC is shown in Table~\ref{tab:dane_auc}.

\begin{table}[htbp]
\centering
\begin{tabular}{cccc}
\toprule
AUC Mean & AUC Std. Dev. & 95\% CI Lower & 95\% CI Upper \\
\midrule
7.876 & 0.336 & 7.218 & 8.548 \\
\bottomrule
\end{tabular}
\caption{Posterior summary of AUC for Dane County outage event.}
\label{tab:dane_auc}
\end{table}

The posterior mean AUC of 7.85, with a 95\% credible interval from 7.22 to 8.55, reflects the cumulative impact of this outage. As a composite metric, AUC captures both the magnitude and duration of the disruption. Compared to other events illustrated in the following section, Dane County falls in the moderate-to-severe range, illustrating the model’s ability to capture real-world variation in outage intensity. This case also demonstrates how spline-based Bayesian inference yields interpretable, uncertainty-aware summaries for operational risk monitoring.

In the next section, we apply the model across a broader set of events, evaluating robustness, calibration, and resilience metrics across heterogeneous counties and outage types.

%%%%%%%%%%%%%%%%%%%%%%%%%%%%%%%%%%%%%%%%%%%%%%%%%%%%%%%%%%%%%
\section{Results}

We apply our hierarchical Bayesian spline model to a curated set of outage events drawn from the EAGLE-I dataset, with a focus on counties across southern Wisconsin. These counties exhibit diverse characteristics—including population density, infrastructure complexity, and weather exposure—making them ideal for evaluating the robustness, flexibility, and interpretability of our model. In this section, we organize the results thematically, using case-based archetypes to demonstrate performance across a spectrum of real-world outage dynamics. We then synthesize findings through cross-event metrics that reflect both methodological behavior and operational insight.

\subsection{Case Studies by Outage Type}

To assess the flexibility and robustness of our hierarchical Bayesian spline model, we apply it to a selection of representative power outage events from southern Wisconsin. These case studies are chosen to reflect a spectrum of outage conditions—ranging from abrupt, short-lived disruptions to extended, low-intensity events—and to probe the model’s capacity to adaptively smooth, quantify uncertainty, and provide interpretable summaries under diverse temporal and geographic contexts. Together, they highlight how posterior behavior responds to differences in event severity, data richness, and system volatility.

\subsubsection{Urban Thunderstorm: Dane County (May 22, 2024)}

A severe thunderstorm caused over 40,000 customer outages in Dane County, with impacts concentrated over a few hours and a relatively clean restoration profile.

\begin{figure}[htbp]
\centering
\includegraphics[width=0.6\textwidth]{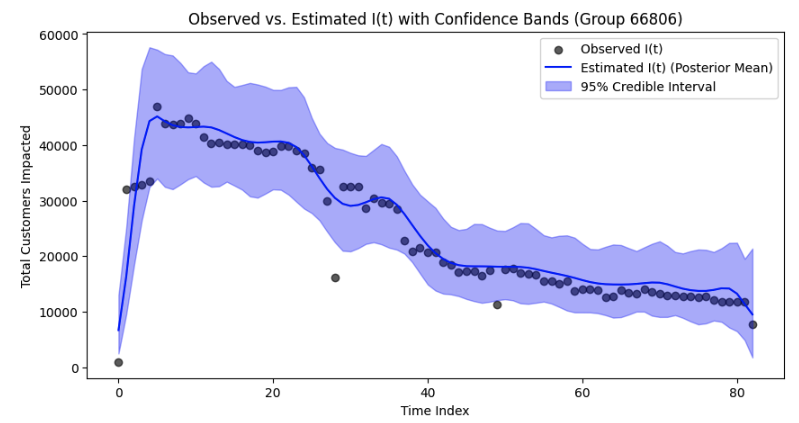}
\caption{Posterior predictive fit for Dane County (May 22, 2024).}
\label{fig:dane_fit}
\end{figure}

The model’s posterior mean tracks the observed trajectory closely, and 95\% credible intervals exhibit strong calibration, with 95.2\% of observations falling within bounds. The estimated area under the curve (AUC) is 7.88 (95\% CI: 7.22--8.55), suggesting moderate cumulative risk. The RMSE of 3346.1 reflects the scale of customer impact but remains well-controlled given the event size and complexity. The LOO-CV score of –849.3 indicates good out-of-sample generalization. This case illustrates the model’s effectiveness in reconstructing high-resolution, well-structured urban outages while maintaining uncertainty quantification appropriate to the data richness.

\subsubsection{Extended Snowstorm: Door County (April 2--3, 2024)}

An early April snowstorm in Door County resulted in persistent outages dispersed over more than 300 time points. Though relatively low in intensity, the event’s duration makes it a valuable test case for temporal resilience modeling.

\begin{figure}[htbp]
\centering
\includegraphics[width=0.6\textwidth]{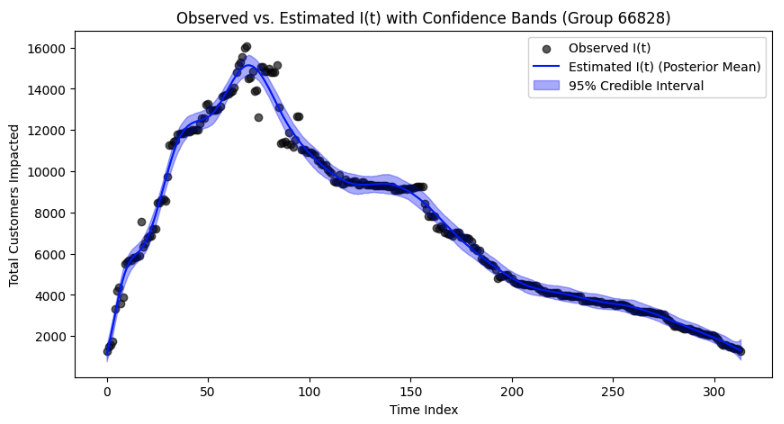}
\caption{Posterior predictive fit for Door County (April 2--3, 2024).}
\label{fig:door_fit}
\end{figure}

The model captures the gradual rise and long-tailed decline in risk, yielding an AUC of 85.53 (95\% CI: 84.37--86.72)—the highest in the dataset. Posterior predictive coverage is 83.8\%, a notable improvement over earlier configurations, though slight underdispersion remains during prolonged low-variance segments. The RMSE is 389.5, reflecting the relatively smooth nature of the outage curve, while the LOO-CV score of –2403.9 indicates model complexity is well-balanced with generalization. Moderate directional bias (123 underpredictions, 191 overpredictions) reflects how spline smoothing averages over frequent but minor fluctuations. Nonetheless, this case demonstrates the model’s capacity to accommodate long-term accumulation of risk while maintaining numerical and inferential stability.

\subsubsection{Urban Blizzard: Milwaukee and Waukesha Counties (January 12--15, 2024)}

A mid-January blizzard led to widespread outages in both Milwaukee and Waukesha Counties, with high winds and heavy snow causing peaks in excess of 60,000 customers affected. The events exhibit steep onset and heterogeneous recovery, challenging the model’s smoothness and variance assumptions.

\begin{figure}[htbp]
\centering
\includegraphics[width=0.45\textwidth]{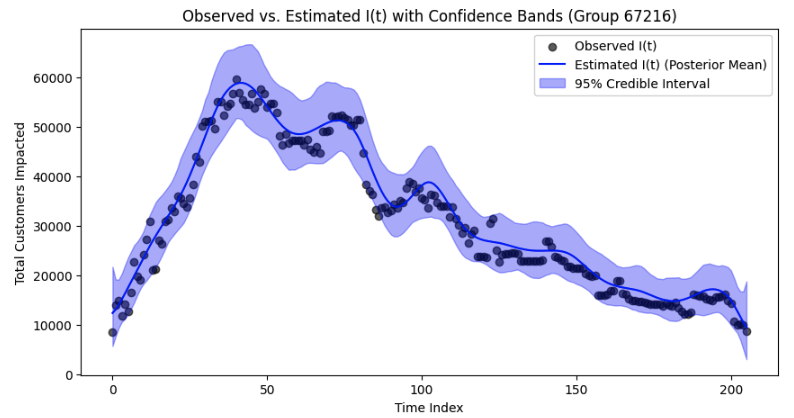}
\includegraphics[width=0.45\textwidth]{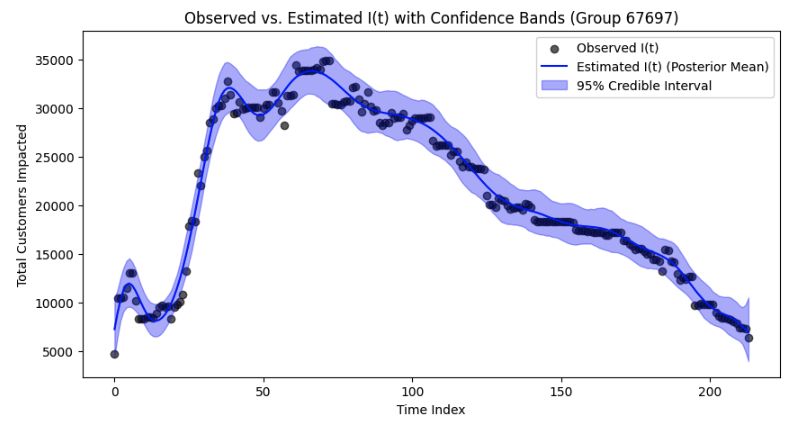}
\caption{Posterior predictive fits for Milwaukee (left) and Waukesha (right) Counties (January 12--15, 2024).}
\label{fig:milwaukee_waukesha_fit}
\end{figure}

Milwaukee’s AUC of 11.06 and Waukesha’s 34.18 reflect the severity and persistence of the January 2024 blizzard. Posterior coverage is high—98.5\% for Milwaukee and 95.8\% for Waukesha—and the RMSEs (2472.9 for Milwaukee, 948.3 for Waukesha) correspond to the scale of affected populations and volatility of the events. Milwaukee’s LOO-CV score of –2089.4 and Waukesha’s –1963.5 indicate the model’s ability to generalize well even under complex dynamics. Nonetheless, both counties exhibit underprediction during peak intensity (Milwaukee: 48 under, 158 over; Waukesha: 82 under, 132 over), a pattern consistent with spline smoothness constraints. These events suggest the need for greater local adaptivity—such as changepoint detection or time-varying dispersion—in future iterations of the model.

\subsubsection{Brief Severe Events: Jefferson County (Multiple Events, 2024)}

Jefferson County experienced a series of brief, severe outages associated with localized wind and storm activity. These events offer insight into the model’s behavior under sparse and rapidly evolving data conditions.

\begin{figure}[htbp]
\centering
\includegraphics[width=0.3\textwidth]{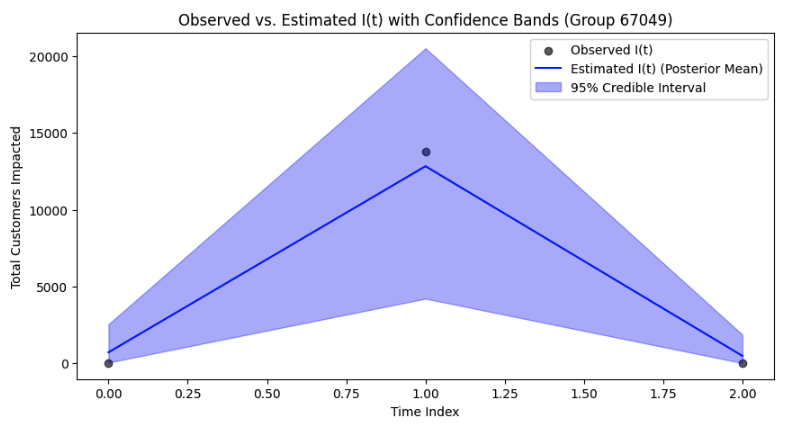}
\includegraphics[width=0.3\textwidth]{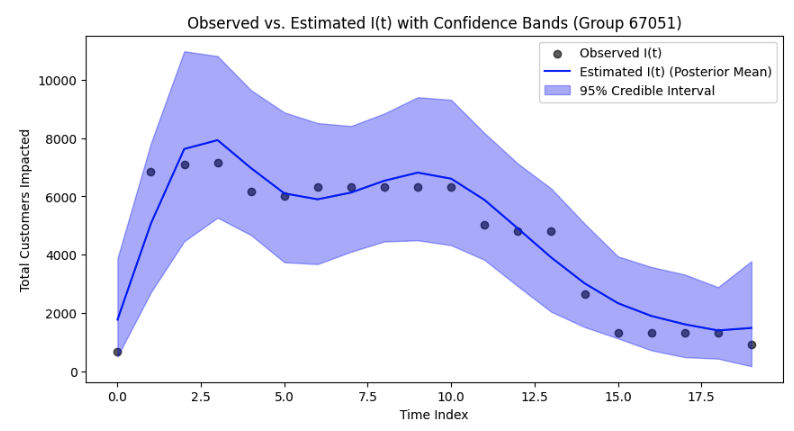}
\includegraphics[width=0.3\textwidth]{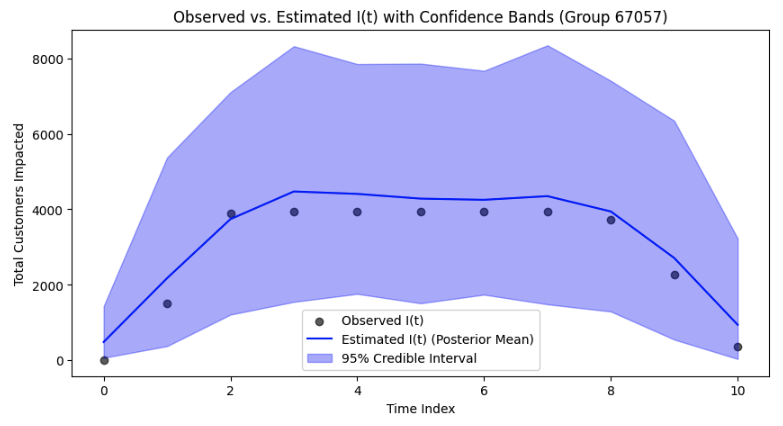} \\
\includegraphics[width=0.3\textwidth]{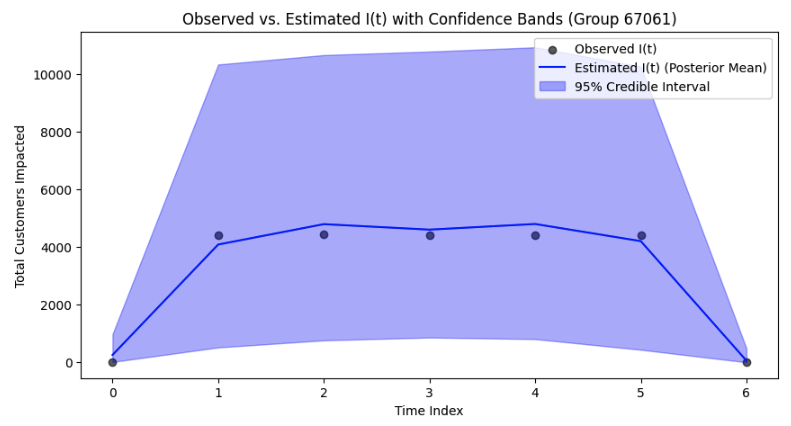}
\caption{Posterior predictive fits for Jefferson County events across 2024.}
\label{fig:jefferson_fits}
\end{figure}

\begin{itemize}
\item \textbf{May 29}: Brief 30-minute outage; AUC 0.55, coverage 33.3\%
\item \textbf{June 23}: Tornado-driven disruption; AUC 3.75, coverage 100\%
\item \textbf{August 6}: Wind and thunderstorm; AUC 1.42, coverage 90.9\%
\item \textbf{October 31}: High winds; AUC 0.92, coverage 85.7\%
\end{itemize}

These events display the model’s ability to manage sparse, low-information regimes with credible intervals that expand appropriately. RMSE values range from 220.8 (Minimal) to 810.6 (Short), reflecting modest reconstruction error given the brevity and volatility of the events. ELPD-LOO scores, such as –22.1 for the 3-point Short event and –52.1 for Minimal, highlight the limited information content and the corresponding challenge of out-of-sample generalization. Despite this, posterior bias remains low across all Jefferson variants (e.g., 1 under and 2 over in Short), and coverage is generally strong—100\% in the Moderate case and above 85\% in the others. In these scenarios, partial pooling and flexible B-spline bases contribute to both numerical stability and plausible uncertainty estimates, even when the observed data provide minimal temporal structure.

\subsection{Cross-Event Posterior Behavior}

While case studies provide detailed illustrations of the model's performance under specific conditions, we now turn to aggregated diagnostics across all events to evaluate broader trends in posterior behavior. This allows us to assess whether the model is systematically well-calibrated, capable of delivering stable uncertainty estimates, and effective at extracting meaningful resilience metrics across a diverse array of outage scenarios.

We organize this analysis into three components: calibration and directional bias; AUC-based resilience scoring; and uncertainty behavior as a function of data richness.

\subsubsection{Posterior Calibration and Directional Bias}

A fundamental criterion for evaluating the reliability of a Bayesian model is posterior predictive calibration—that is, whether the credible intervals constructed from the posterior predictive distribution actually reflect the variability in the observed data. In Table~\ref{tab:pp_coverage}, we report the proportion of observed time points that fall within the model’s 95\% posterior predictive intervals for each event.

\begin{table}[htbp]
\centering
\caption{Posterior predictive 95\% coverage and credible interval width.}
\label{tab:pp_coverage}
\begin{tabular}{lcccc}
\toprule
\textbf{County (Group)} & \textbf{Obs.} & \textbf{Coverage} & \textbf{Misses} & \textbf{Avg CI Width} \\
\midrule
Jefferson Moderate (67051) & 20 & 1.000 & 0 & 4190 \\
Jefferson Minimal (67061) & 7 & 0.857 & 1 & 7373 \\
Jefferson Brief (67057) & 11 & 0.909 & 1 & 5373 \\
Dane (66806) & 83 & 0.952 & 4 & 15805 \\
Waukesha (67697) & 214 & 0.958 & 9 & 4272 \\
Milwaukee (67216) & 206 & 0.985 & 3 & 11662 \\
Door (66828) & 314 & 0.838 & 51 & 776 \\
Jefferson Short (67049) & 3 & 0.333 & 2 & 16506 \\
\bottomrule
\end{tabular}
\end{table}

Events such as Jefferson Moderate and Dane County show excellent calibration, with 100\% and 95.2\% of observed data points, respectively, falling within the 95\% posterior predictive intervals. Similarly, Milwaukee and Waukesha show high coverage rates of 98.5\% and 95.8\%, respectively, suggesting that the model is able to represent uncertainty adequately in large-scale, high-resolution urban events. In contrast, coverage drops in the most temporally extended case—Door County—where the model captures only 83.8\% of observations. The lowest coverage is seen in Jefferson Short (33.3\%), which includes only three time points and illustrates the inherent volatility of extremely sparse data. These results indicate that while the model calibrates well across a range of structured and high-density cases, it may still slightly underrepresent uncertainty in prolonged low-magnitude events or in ultra-sparse trajectories, pointing to opportunities for refined variance adaptation in these extremes.

We also examine directional bias—that is, whether the posterior predictive mean tends to systematically under- or overestimate the observed data. Table~\ref{tab:pp_bias} presents the number of underpredicted and overpredicted time points per group.

\begin{table}[htbp]
\centering
\caption{Directional bias in posterior predictive means.}
\label{tab:pp_bias}
\begin{tabular}{lcc}
\toprule
\textbf{County (Group)} & \textbf{Underpredicted Points} & \textbf{Overpredicted Points} \\
\midrule
Door (66828) & 123 & 191 \\
Milwaukee (67216) & 48 & 158 \\
Waukesha (67697) & 82 & 132 \\
Dane (66806) & 16 & 67 \\
Jefferson Moderate (67051) & 4 & 16 \\
Jefferson Brief (67057) & 1 & 10 \\
Jefferson Minimal (67061) & 2 & 5 \\
Jefferson Short (67049) & 1 & 2 \\
\bottomrule
\end{tabular}
\end{table}

Directional bias patterns remain consistent with expectations. Milwaukee (48 underpredictions, 158 overpredictions) and Waukesha (82 underpredictions, 132 overpredictions) both exhibit a predominance of overprediction, likely due to extended recovery tails following steep outage peaks. These results suggest that the model’s smoothing may slightly favor persistence over abrupt correction—an effect of regularization and basis adaptation. Door County shows the most balanced directional behavior (123 underpredictions, 191 overpredictions), consistent with its gradual, flat profile. Meanwhile, brief events like Jefferson Minimal, Short, and Brief show minimal directional bias (typically one or two off-target points), suggesting that the model is not overly reactive in sparse or short-lived cases. This balance between responsiveness and stability illustrates the spline model’s ability to generalize across temporal resolutions, albeit with room for improvement near sharp inflections.

\subsubsection{AUC Distributions as Resilience Metrics}

To translate outage trajectories into actionable resilience summaries, we compute the area under the posterior mean risk curve (AUC) for each event. This scalar quantity captures both the severity (height) and duration (length) of the outage, enabling cross-event comparisons.

\begin{table}[htbp]
\centering
\caption{Posterior summaries of AUC (Area Under the Curve) for selected outage events.}
\label{tab:auc_summary}
\begin{tabular}{lrrr}
\toprule
\textbf{County (Group)} & \textbf{AUC Mean} & \textbf{AUC Std} & \textbf{95\% CI} \\
\midrule
Dane (66806) & 7.88 & 0.32 & $(7.25, 8.52)$ \\
Door (66828) & 85.53 & 0.51 & $(84.55, 86.52)$ \\
Jefferson Short (67049) & 0.54 & 0.27 & $(0.14,1.16)$ \\
Jefferson Moderate (67051) & 3.73 & 0.37 & $(3.06,4.50)$ \\
Jefferson Brief (67057) & 1.41 & 0.28 & $(0.91, 2.03)$ \\
Jefferson Minimal (67061) & 0.90 & 0.26 & $(0.44, 1.46)$ \\
Milwaukee (67216) & 11.05 & 0.23 & $(10.62, 11.50)$ \\
Waukesha (67697) & 34.18 & 0.40 & $(33.43, 34.99)$ \\
\bottomrule
\end{tabular}
\end{table}

AUC values align well with the temporal and structural profiles of each outage event. Door County registers the highest AUC (85.53), consistent with its long, low-severity snowstorm and persistent service interruptions. Waukesha (34.18) and Milwaukee (11.06) also exhibit substantial cumulative risk, reflecting the magnitude and slow resolution of the January blizzard. Dane County’s moderate AUC (7.88) corresponds to a short, well-managed thunderstorm with rapid recovery. In contrast, Jefferson County events remain low in cumulative impact, with AUCs ranging from 0.55 (Short) to 3.75 (Moderate), underscoring their brief and localized nature. Notably, the standard deviations across AUC estimates remain narrow—even in sparse regimes—demonstrating the model’s ability to deliver stable posterior summaries while adapting to heterogeneous event profiles.

\subsection{Regional Resilience Comparison}

The case studies above illustrate the model's flexibility across diverse outage types, but its true utility lies in how posterior summaries can support region-wide resilience assessment. By comparing events across counties, we identify consistent patterns in model behavior and risk expression:

\begin{itemize}
\item \textbf{Dane County}: A well-monitored urban thunderstorm with rapid recovery, yielding a moderate AUC and strong calibration.
\item \textbf{Door County}: A long, low-severity snowstorm resulting in the highest AUC due to duration; coverage is modest, reflecting minor variance misspecification.
\item \textbf{Milwaukee and Waukesha Counties}: High-density blizzard events marked by steep spikes and tailing recoveries; both show high cumulative impact and directional underprediction.
\item \textbf{Jefferson County}: Brief, localized disruptions with minimal cumulative risk; model uncertainty expands appropriately in sparse-data regimes, aided by partial pooling.
\end{itemize}

Taken together, these comparisons underscore the model’s robustness to data richness, event duration, and spatial heterogeneity. The posterior outputs—especially AUC distributions and credible intervals—provide interpretable, event-specific summaries that can inform regional grid resilience planning. This positions the model as a practical tool for benchmarking performance, evaluating recovery dynamics, and informing targeted infrastructure investments.

\subsection{Comparison with Naive Trapezoidal AUC}
\label{sec:naive_comparison}

To assess the added value of posterior AUC estimation relative to a deterministic 
baseline, we compare our Bayesian estimates against a naive trapezoidal AUC computed 
directly from raw observed proportions $\hat{p}_{g,t} = y_{g,t} / n_{g,t}$, without 
any model, smoothing, or uncertainty quantification. This baseline approximates the 
approach implicitly taken by prior resilience frameworks that compute cumulative outage 
metrics directly from observed time series \cite{lee2024quantifying, 
carrington2021extracting}.

Table~\ref{tab:auc_comparison} presents the naive AUC alongside the posterior mean 
AUC, 95\% credible intervals, and divergence expressed in posterior standard deviation 
units for each event. Across all eight events, the naive estimate falls within the 
posterior 95\% credible interval, with divergence ranging from 0.01$\sigma$ (Jefferson 
Short, $T=3$) to 1.88$\sigma$ (Milwaukee, $T=206$). This consistent agreement on 
point estimates serves as a validation result: the Bayesian model recovers the same 
central tendency as the naive integrator while adding principled uncertainty 
quantification.

\begin{table}[htbp]
\centering
\caption{Comparison of naive trapezoidal AUC against posterior Bayesian AUC.
Divergence is expressed in units of posterior standard deviations. All naive estimates
fall within the posterior 95\% credible interval, validating the posterior mean while
demonstrating that the naive approach provides no characterization of estimate
reliability.}
\label{tab:auc_comparison}
\small
\begin{tabular}{lrrrrrr}
\toprule
\textbf{County (Group)} & \textbf{$T$} & \textbf{Naive} &
\textbf{P.\ ($\mu$)} & \textbf{CI Low} & \textbf{CI High} &
\textbf{Div.\ ($\sigma$)} \\
\midrule
Jefferson Short (67049)    & 3   & 0.556  & 0.552  & 0.138  & 1.166  & 0.01 \\
Jefferson Minimal (67061)  & 7   & 0.889  & 0.911  & 0.475  & 1.461  & 0.09 \\
Jefferson Brief (67057)    & 11  & 1.259  & 1.391  & 0.915  & 1.948  & 0.49 \\
Jefferson Moderate (67051) & 20  & 3.550  & 3.714  & 3.029  & 4.480  & 0.44 \\
Dane (66806)               & 83  & 7.512  & 7.876  & 7.237  & 8.555  & 1.09 \\
Milwaukee (67216)          & 206 & 10.602 & 11.046 & 10.600 & 11.517 & 1.88 \\
Waukesha (67697)           & 214 & 33.847 & 34.176 & 33.415 & 34.974 & 0.82 \\
Door (66828)               & 314 & 85.252 & 85.523 & 84.514 & 86.536 & 0.54 \\
\bottomrule
\end{tabular}
\end{table}

The critical distinction between the two approaches lies not in the point estimates but in what each method communicates about the reliability of those estimates. The naive AUC for Jefferson Short ($T=3$) is 0.556, and the naive AUC for Door County ($T=314$) is 85.252. Both are scalar values with no attached uncertainty — the naive framework treats them as equally well-characterized. The posterior credible intervals tell a fundamentally different story: the Jefferson Short interval spans $(0.14, 1.17)$, a range nearly an order of magnitude wide relative to its mean, while the Door County interval spans $(84.51, 86.54)$, a range of approximately 2.0 units. A decision-maker relying on naive AUC alone has no basis for distinguishing that the Jefferson Short estimate is highly uncertain while the Door County estimate is precise. The posterior credible interval makes this distinction explicit and actionable.

This pattern has direct operational consequences. When comparing recovery performance across two counties, overlapping credible intervals indicate that the difference in AUC is not statistically meaningful given the available data — a conclusion that cannot be drawn from point estimates alone. Similarly, when ranking events by severity for resource allocation or regulatory reporting, the width of the credible interval determines whether that ranking is reliable or potentially an artifact of sparse observation. The naive approach produces rankings without any indication of which comparisons are trustworthy, whereas the posterior framework surfaces this uncertainty as a first-class output.

In summary, the comparison confirms that our Bayesian model recovers AUC point estimates consistent with the naive baseline across all data regimes, while providing the uncertainty quantification that prior deterministic approaches structurally cannot. The value of the proposed framework is therefore not in correcting the naive point estimate, but in characterizing how much that estimate should be trusted — a distinction that is especially consequential in the sparse, heterogeneous data conditions typical of real-world power outage monitoring.

In the following section, we explore methodological implications, modeling assumptions, and opportunities for future development.

%%%%%%%%%%%%%%%%%%%%%%%%%%%%%%%%%%%%%%%%%%%%%%%%%%%%%%%%%%%%%
\section{Discussion}

This study develops and applies a hierarchical Bayesian spline model to estimate time-varying outage risk from customer impact data. Leveraging a Beta-Binomial likelihood with group-specific spline trajectories and partial pooling over dispersion, the model balances flexibility and interpretability while retaining computational tractability. Our evaluation across diverse power outage events in southern Wisconsin demonstrates the model’s ability to accommodate heterogeneity in event scale, duration, and structure—ranging from brief tornadoes to prolonged snowstorms—while yielding posterior summaries that inform both technical diagnostics and real-world resilience decisions.

\subsection{Modeling Contributions}

The proposed framework offers several methodological and applied innovations spanning spline-based flexibility, uncertainty-aware inference, partial pooling for dispersion stabilization, and interpretability with application readiness.

A central empirical finding — confirmed by direct comparison against a naive trapezoidal baseline in Section~\ref{sec:naive_comparison} — is that the posterior mean AUC is consistent with deterministic integration of raw observed proportions across all eight events studied. Naive and posterior estimates agree within 1.88 posterior standard deviations in every case, with the naive estimate falling inside the 95\% credible interval universally. This agreement validates the posterior mean as a reliable point estimate, while simultaneously exposing the core insufficiency of deterministic approaches: because the naive method produces a scalar with no uncertainty attached, it is structurally incapable of communicating whether that estimate is well-supported or highly uncertain. The posterior credible interval width varies from a range of approximately 1.0 units for Jefferson Short ($T=3$) to 2.0 units for Door County ($T=314$), encoding data richness directly into the resilience summary in a way that deterministic methods cannot.

For spline-based flexibility, a fixed-degree B-spline basis enables smooth estimation of time-varying probabilities without imposing strict parametric forms. This permits the model to capture heterogeneous dynamics—from sharp urban disruptions to slow rural recovery trajectories—using a unified functional framework. Uncertainty-aware inference is obtained as posterior distributions reflect both observation density and event volatility, supporting well-calibrated credible intervals and resilience metrics such as AUC and RMSE. This contrasts with point-estimate models that often mask uncertainty in sparse or rapidly evolving settings.

By hierarchically modeling the dispersion parameter across groups, the model borrows strength in low-data regimes (e.g., brief events in Jefferson County) while preserving fidelity in data-rich counties (e.g., Milwaukee, Door). This partial pooling for dispersion stabilization stabilizes estimation without introducing artificial shrinkage where local information dominates. Interpretability is further supported by posterior predictive intervals, AUC estimates, and directional bias diagnostics, which together provide transparent, decision-relevant summaries. As a part of application readiness, these outputs can be directly used by grid operators to benchmark performance, identify recovery bottlenecks, and inform infrastructure investments.

\subsection{Future Directions}

Despite strong empirical performance, several limitations remain. Limitations indicated here focus on rigid constraints such as a fixed basis structure and uniform temporal spacing. Assumptions made on information within the model, such as a lack of covariate information and spatial structure, may yield wider posterior variance. 

Under a fixed-basis structure, the use of degree-3 B-splines with fixed knots imposes a smoothness constraint that may limit responsiveness to sudden transitions, double peaks, or step-like recoveries. Event-specific or adaptive bases could improve flexibility without sacrificing interpretability. Similarly, the model assumes regularly spaced time indices, which may not reflect the reporting cadence of real-world outage data sets. Handling irregular or missing time points will require extensions to latent time models or continuous-time interpolation.

The current formulation treats group-level dynamics as exchangeable beyond the hierarchical structure. Incorporating exogenous covariates such as population density, vegetation cover, wind speed, or repair delay—could enhance model accuracy and support explanatory insights. Aside from these potential covariate improvements, spatial relationships of the power grid may improve posterior estimation. While information is partially shared across counties through the hierarchical prior on dispersion, the model treats geographic groups as exchangeable and lacks spatial priors or adjacency-aware pooling. Recent spatiotemporal frameworks \citep{xu2024hazard, jiang2024spatiotemporal} demonstrate the value of propagating risk across connected network components; incorporating analogous spatial structure---through CAR models, spatial Gaussian processes, or graph-based pooling---could further stabilize inference in under-observed areas and enable recovery trajectory estimation at the network level.

Building on this foundation, several extensions offer paths forward:

\begin{itemize}
    \item \textbf{Changepoint or regime-switching components} could allow the model to adapt to abrupt transitions or intervention effects, such as repair crew arrival or grid reconfiguration, that are not well captured by globally smooth trajectories.
    
    \item \textbf{Joint modeling with exogenous data sources} (e.g., storm tracks, vegetation encroachment, infrastructure age) could improve forecast accuracy and causal interpretation, transforming the model from a passive observer to a proactive diagnostic tool.
    
    \item \textbf{Spatial extensions} via structured priors or neighbor-based shrinkage would enable the model to propagate information across geographic boundaries—especially useful when one region has dense data and another is sparsely reported.
    
    \item \textbf{Latent class or clustering structures} could help identify common outage archetypes across counties or years, aiding operational planning by categorizing events into high-impact vs. nuisance outages or slow-recovery vs. fast-recovery profiles.

    \item \textbf{Minimum observation thresholds or informative priors} for ultra-sparse events ($T \leq 5$) could improve calibration in regimes where the likelihood provides minimal information. Events such as Jefferson Short ($T = 3$) suggest that practitioner-informed priors on recovery shape or duration may be necessary to yield well-calibrated intervals when data are extremely limited.
    
\end{itemize}

\subsection{Conclusion}

Hierarchical Bayesian spline models offer a robust and interpretable framework for analyzing time-varying power outage risk. Our results show that the posterior distributions not only track observed disruptions but also provide meaningful uncertainty quantification, particularly when data are sparse or dynamics are volatile. Comparison against a naive trapezoidal baseline confirms that the posterior mean AUC is consistent with deterministic integration across all data regimes, while the posterior credible interval provides the reliability characterization that deterministic approaches structurally cannot. In an era of increasing climate-driven hazards, such models offer utilities and emergency planners a principled method for understanding, comparing, and preparing for grid failure events — one that makes explicit not just what the data show, but how much confidence those summaries warrant. The flexibility of the Bayesian approach, combined with computational tractability and interpretability, positions it as a critical component in the evolving toolkit of infrastructure resilience analytics.

%%%%%%%%%%%%%%%%%%%%%%%%%%%%%%%%%%%%%%%%%%%%%%%%%%%%%%%%%%%%%
\section*{Acknowledgments}
This material is based upon work supported by the U.S. Department of Energy, Office of Cybersecurity, Energy Security, and Emergency Response (CESER), under contract number DE-AC05-00OR22725. The authors also acknowledge the support of the CESER for the development and maintenance of the EAGLE-I\texttrademark\ platform, which provided the power outage data used in this study.

%%%%%%%%%%%%%%%%%%%%%%%%%%%%%%%%%%%%%%%%%%%%%%%%%%%%%%%%%%%%%

% Credit authorship
\printcredits

%% Bibliography
\bibliographystyle{cas-model2-names}
\bibliography{bayesian_outages_refs}

\end{document}